\documentclass[conference]{IEEEtran}
\usepackage{amssymb,relsize,graphicx,amsmath,psfig,cite,float,setspace,braket}
\usepackage[left=0.680in,right=0.673in,top=0.860in,bottom=1.07in]{geometry}
\usepackage{amssymb,relsize,graphicx,amsmath,psfig,cite,float,subfigure,setspace}
\usepackage{xcolor}
\usepackage{tabularx}
\usepackage{xcolor}
\usepackage{tabularx}
\usepackage{flushend}
\usepackage{algorithm}
\usepackage{algpseudocode}
\usepackage{subcaption}
\usepackage{subfigure}
\usepackage{bm}
\usepackage{algorithm}
\usepackage{amsmath}
\algnewcommand\algorithmicinput{\textbf{Input:}}
\algnewcommand\algorithmicinit{\textbf{Init:}}
\algnewcommand\algorithmicoutput{\textbf{Output:}}
\algnewcommand\INPUT{\item[\algorithmicinput]}
\algnewcommand\Init{\item[\algorithmicinit]}
\algnewcommand\OUTPUT{\item[\algorithmicoutput]}

\newcommand{\bi}{\begin{itemize}}
\newcommand{\ei}{\end{itemize}}

\newcommand{\be}{\begin{enumerate}}
\newcommand{\ee}{\end{enumerate}}
\newcommand{\bd}{\begin{description}}
\newcommand{\ed}{\end{description}}
\newcommand{\bc}{\begin{center}}
\newcommand{\ec}{\end{center}}
\newcommand{\bt}{\begin{tabbing}}
\newcommand{\et}{\end{tabbing}}
\newcommand{\bfig}{\begin{figure}}
\newcommand{\efig}{\end{figure}}
\newcommand{\beq}{\begin{equation}}
\newcommand{\beqarr}{\begin{eqnarray}}
\newcommand{\beqarrn}{\begin{eqnarray*}}
\newcommand{\eeq}{\end{equation}}
\newcommand{\eeqarr}{\end{eqnarray}}
\newcommand{\eeqarrn}{\end{eqnarray*}}
\newcommand{\bflr}{\begin{flushright}\vspace{-0.2in}}
\newcommand{\eflr}{\end{flushright}}
\newcommand{\bsub}{\begin{subequations}}
\newcommand{\esub}{\end{subequations}}
\newcommand{\barr}{\begin{array}}
\newcommand{\earr}{\end{array}}
\newcommand{\nn}{\nonumber}

\def\undb#1{\mbox{\bf{#1}}}

\def\dn{\stackrel{\scriptscriptstyle \triangle}{=}}

\def\BibTeX{{\rm B\kern-.05em{\sc i\kern-.025em b}\kern-.08em
		T\kern-.1667em\lower.7ex\hbox{E}\kern-.125emX}}

\begin{document}

\title{\huge{Symbol Error Analysis for Fluid Antenna Systems\\ with One- and Two-Dimensional Modulation Schemes} \vspace{-0.7cm}}

\author{\IEEEauthorblockN{Soumya P. Dash\IEEEauthorrefmark{1} and George C. Alexandropoulos\IEEEauthorrefmark{2}}
\IEEEauthorblockA{\IEEEauthorrefmark{1}School of Electrical and Computer Sciences, Indian Institute of Technology Bhubaneswar, Khordha, Odisha, India \\
\IEEEauthorrefmark{2}Department of Informatics and Telecommunications, National and Kapodistrian University of Athens, 16122 Athens, Greece \\
{e-mails: spdash@iitbbs.ac.in, alexandg@di.uoa.gr}}
\vspace{-0.9cm}
}

{}
\maketitle
\begin{abstract}
This paper considers a Fluid Antenna (FA) system comprising a single-antenna transmitter that communicates with a receiver equipped with an FA array with $N$ ports. The transmitter is assumed to deploy any of the modulation schemes: \textit{i}) two-sided $M$-ary amplitude-shift keying, \textit{ii}) $M$-ary phase-shift keying, iii) $M$-ary quadrature-amplitude modulation, and \textit{iv}) binary frequency-shift keying, the channels between its antenna and the receiver ports are subjected to Rayleigh fading, and the receiver chooses the best $K$ out of its $N$ ports for symbol detection. Considering that the receiver combines the signals from the best $K$ ports using maximal-ratio combining, the optimal reception structures for all the considered signaling schemes are obtained. We also present novel exact closed-form expressions for the respective symbol error probabilities (SEPs) of the FA system, as well as asymptotic approximations valid at high signal-to-noise ratios. The presented analysis is corroborated through comparisons with simulation results, showcasing the critical role of various system parameters on the SEP performance.
\end{abstract}
\begin{IEEEkeywords}
Fluid antenna system, one- and two-dimensional modulation schemes, port correlation, symbol error probability.
\end{IEEEkeywords}
\section{Introduction}
Multiple-Input Multiple-Output (MIMO) has been a cornerstone of wireless communications since the advent of fourth-generation systems. In fifth-generation and beyond networks, this paradigm has evolved into massive MIMO (mMIMO) to support demanding use cases, such as massive machine-type communications, ultra-reliable low-latency communications, and enhanced mobile broadband \cite{10379539}. However, scaling antenna arrays to realize mMIMO systems incurs significant challenges, including increased power consumption and hardware complexity due to the large number of required radio frequency chains. Furthermore, the need to stringently maintain half-wavelength antenna spacing imposes practical limitations on array size, particularly in space-constrained devices \cite{11098630}.

Over recent years, researchers have explored reconfigurable antenna placement as an alternative to MIMO, among which Fluid Antenna (FA) systems have emerged as a promising technology~\cite{9264694}. FA systems employ liquid-based or reconfigurable pixel antennas whose locations can be dynamically reconfigured across a set of predefined ports within a given region to optimize system performance. This adaptability provides FA systems with additional spatial degrees of freedom, enabling improved communication performance while maintaining a relatively low overall hardware complexity. Thus, several studies have been reported to demonstrate the performance of FA-assisted wireless systems over the years. For instance, the works in \cite{11098630, 10375698, ganesh2025OPMRC, ganeshgc25} focus on the outage probability of FA system, while their achievable rate is studied in~\cite{10960696, 10934056, 10303274} and their secrecy performance in~\cite{10694739, 10468625, 10978677}. However, to the best of our knowledge, the Symbol Error Probability (SEP) of FA systems still remains an unexplored area, except for the work in \cite{11360610} where the authors study the SEP performance of a FA system selecting the best port for data demodulation.

Motivated by this research gap, and targeting on a more generalized port selection scheme, this paper studies the SEP performance of a system consisting of an $N$-port FA-equipped receiver, of which the best $K\leq N$ ports are selected and combined using Maximal-Ratio Combining (MRC) for data demodulation. The specific contributions of this work are summarized as follows. Considering the transmission of one- and two-dimensional memoryless signaling schemes, namely, two-sided $M$-ary Amplitude-Shift Keying ($M$-ASK), $M$-ary Phase-Shift Keying ($M$-PSK), $M$-ary Quadrature-Amplitude Modulation ($M$-QAM), and Binary Frequency-Shift Keying (BFSK), optimal Maximum Likelihood (ML) receiver structures for the considered FA system are presented. Using a characteristic function (c.f.) approach, novel exact closed-form expressions for the SEP are derived for all considered modulations. In addition, asymptotic expressions for the SEPs at high average signal-to-noise ratio (SNR) levels are presented, showcasing that the FAS achieves a diversity order of $N$. and numerical results are presented to corroborate the analysis.

{\em Notations:} $(\cdot)^T$ and $\lVert \cdot \rVert$ represent the transpose and $\ell_2$-norm operators, respectively. A complex Gaussian random vector with mean $\bm{\mu}$ and covariance $\undb{K}$ is denoted by $\mathcal{CN}(\bm{\mu}, \undb{K})$, while a real Gaussian random variable with mean $\mu$ and variance $\sigma^2$ is denoted by $\mathcal{N}(\mu, \sigma^2)$. The operators $\mathcal{R}\{\cdot\}$ and $\mathcal{I}\{\cdot\}$ extract the real and imaginary parts, respectively, $(\cdot)^*$ denotes complex conjugation, and $\jmath\triangleq\sqrt{-1}$. The expectation operator is given by $\mathbb{E}[\cdot]$, $\undb{0}_{N\times 1}$ denotes the $N\times 1$ zero vector, and $\undb{I}_N$ is the $N\times N$ identity matrix. Furthermore, $Q(\cdot)$ denotes the Gaussian $Q$-function, ${}_1F_2(\cdot;\cdot;\cdot;\cdot)$ the generalized hypergeometric function, $J_1(\cdot)$ the first-order Bessel function of the first kind, and $B_{1/2}(\cdot,\cdot)$ the incomplete Bessel function.
\section{System Model}
We consider an FA system where the transmitter is equipped with a single antenna and the receiver employs an one-dimensional FA with $N$ ports evenly distributed over a linear length of $W \lambda$, where $\lambda$ is the wavelength. Considering that the transmitter transmits a symbol $s$ in a flat fading wireless communication scenario, the complex baseband received signal at the $k$-th FA's port ($k=1,\ldots,N$) is expressed as follows:
\beq
r_k \triangleq h_k s + n_k,
\label{eq1}
\eeq
where $\undb{n} \triangleq \left[ n_1,\ldots,n_N \right]^T \sim {\mathcal{CN}} \left( \undb{0}_{N \times 1}, \sigma_n^2 \undb{I}_N \right)$ is the additive noise vector and $\undb{h} \triangleq \left[ h_1,\ldots,h_N \right]^T \sim {\mathcal{CN}} \left( \undb{0}_{N \times 1}, \undb{K}_h \right)$ is the fading gain vector, with the elements of $\undb{K}_h$ given by:
\beq
\left( \undb{K} \right)_{ij}
\triangleq \left\{ \begin{array}{ll}
\! \! \sigma_h^2 \, , & i=j \\
\! \! \mu^2 \sigma_h^2 \, , & i \neq j 
\end{array} \right.\quad \forall i,j \in \left\{1,\ldots,N \right\},
\label{eq2}
\eeq
where the port correlation coefficient $\mu$ is obtained as \cite{11360610}:
\beq
\mu = \sqrt{2 \left[ {}_1F_2 \left( \frac{1}{2} ; 1 ;
\frac{3}{2} ; -\pi^2 W^2 \right)
- \frac{J_1 \left(2 \pi W \right)}{2 \pi W} \right]}.
\label{eq3}
\eeq
Further, we consider that the symbol $s$ belongs to a set of equiprobable symbols selected from memoryless one-/two-dimensional signaling schemes, namely (i) two-sided $M$-ASK, (ii) $M$-PSK, (iii) $M$-QAM, or (iv) BFSK, implying that:
\beq
s \in \left\{ \begin{array}{ll}
& \! \! \! \! \! \! \! \! \!
\frac{\sqrt{3 E_{\text{av}}}\left( 2m - 1 - M \right)}
{\sqrt{M^2-1}} \, , \,
m=1\ldots, M \, \ \text{($M$-ASK)} \\
& \! \! \! \! \! \! \! \! \!
\sqrt{E_{\text{av}}}
\exp \left\{ \jmath 2 \pi \frac{\left(m-1 \right)}{M} \right\} ,
m=1\ldots, M \,\, \text{($M$-PSK)} \\
& \! \! \! \! \! \! \! \! \!
\frac{\sqrt{3 E_{\text{av}}} \left( 2 m_1 - 1 - \sqrt{M} \right)}
{\sqrt{2 \left(M-1 \right)}}
+ \jmath \frac{\sqrt{3 E_{\text{av}}} \left( 2 m_2 - 1 - \sqrt{M} \right)}
{\sqrt{2 \left(M-1 \right)}} \, , \\
& \! \! \! \! \! \! \! \! \!
m_1 = 1,\ldots, \sqrt{M} \, ,
m_2 = 1,\ldots, \sqrt{M} \, \text{($M$-QAM)} \\
& \! \! \! \! \! \! \! \! \!
\left\{ \sqrt{E_{\text{av}}}, \jmath \sqrt{E_{\text{av}}} \right\} \, \ \text{(BFSK)} \, 
\end{array}
\right.,
\label{eq4}
\eeq
where $E_{\text{av}}$ denotes the average energy.

Let us define the instantaneous received SNR at the $k$-th port by $\gamma_k \dn \frac{A^2 \left|h_k \right|^2}{\sigma_n^2}$, and let $\gamma_{[1]}, \gamma_{[2]}, \ldots, \gamma_{[N]}$ denote the instantaneous SNRs in descending order, i.e., that $\gamma_{[1]} > \gamma_{[2]} > \ldots > \gamma_{[N]}$. Considering that the FA receiver chooses the best $K$ of the $M$ ports (the $K$ ports with the highest instantaneous SNRs) and combines them using the MRC technique, the instantaneous SNR at the receiver is defined as follows:
\beq
\gamma_{\text{FAS}} \triangleq \sum_{k=1}^K \gamma_{[k]}
= \frac{E_{\text{av}} \left \lVert \undb{h}_{[K]} \right \rVert^2}{\sigma_n^2} \, ,
\label{eq5}
\eeq
where $\undb{h}_{[K]} \triangleq \left[ h_{[1]},\ldots, h_{[K]} \right]^T$ denotes the vector of the channel gains corresponding to the best $K$ instantaneous SNRs. Analytical results for a hybrid selection/MRC in uniformly correlated Nakagami-$m$ faded channels are presented in \cite{1025509}. For the Rayleigh case (i.e., when setting the Nakagami shape parameter as $m=1$) and the number of selected receive branches set as $K$ in \cite[eq. (38a)]{1025509}, the c.f. of $\gamma_{\text{FAS}}$, $\Psi_{\gamma_{\text{FAS}}} \left( \jmath \omega \right)
\triangleq \mathbb{E} \left[ e^{\jmath \omega \gamma_{\text{FAS}}} \right]$, is obtained as follows:
\begin{small}
\begin{align}
\! \! \! \! \Psi_{\gamma_{\text{FAS}}} \left( \jmath \omega \right)
=& \left( \frac{1 - \mu^2}{1+\left(N-1 \right) \mu^2} \right)
\sum_{p=0}^{\infty} \left( \frac{\mu^2}
{1+\left(N-1 \right) \mu^2} \right)^p \nn \\
& \times \hspace{-0.5cm} \sum_{\begin{array}{c}
{\scriptstyle \left(\ell_1,\ldots, \ell_N \right)} \\
{\scriptstyle 0 \leq \ell_1,\ldots, \ell_N \leq p ;} \\
{\scriptstyle \ell_1 + \ell_2 + \ldots + \ell_N = p}
\end{array}} \hspace{-0.5cm}
{p \choose \ell_1,\ldots,\ell_N} \,
{\mathcal{I}}_{p,\ell_1,\ldots,\ell_N}
\left( \jmath \omega \right) \, ,
\label{eq6}
\end{align}
\end{small}
\hspace{-0.12cm}where ${p \choose \ell_1,\ldots,\ell_N} = \frac{p!}{\prod_{k=1}^N \ell_k!}$ and:
\begin{small}
\begin{align}
& \! \! \! \! {\mathcal{I}}_{p,\ell_1,\ldots,\ell_N}
\left( \jmath \omega \right) \nn \\
&\triangleq \hspace{-0.8cm} \sum_{\begin{array}{c}
{\scriptstyle \left( q_1 , \ldots, q_N \right)} \\
{\scriptstyle q_1,\ldots,q_N \geq 0 ;} \\
{\scriptstyle \sum_{i=1}^k q_i \leq \sum_{i=1}^k \ell_i ,} \\
{\scriptstyle k=1,\ldots, N-1 ;} \\
{\scriptstyle \sum_{k=1}^N q_k = p}
\end{array}} \hspace{-0.2cm}
\frac{\prod\limits_{k=2}^N \frac{1}{k^{q_k}}
{\ell_k + \sum_{j=1}^{k-1} \left(\ell_j - q_j \right) \choose \ell_k}}
{\left[\begin{array}{c}
\! \! \! \prod\limits_{k=1}^K \left(1 - \jmath \omega
\left(1-\mu^2 \right) \Gamma_{\text{av}} \right)^{q_k+1} \\
\! \! \! \times \prod\limits_{k=K+1}^N \left( 1 - \jmath \omega
\frac{\left( 1-\mu^2 \right) K \Gamma_{\text{av}}}{k} \right)^{q_k+1}
\end{array} \! \! \! \right]} ,
\label{eq7}
\end{align}
\end{small}
\hspace{-0.12cm}with $\Gamma_{\text{av}} \triangleq \frac{E_{\text{av}} \sigma_h^2}{\sigma_n^2}$ being the average SNR of the FA system.

Let $r_{[1]},\ldots, r_{[N]}$ denote the first $K$ largest-SNR received signals at the FA-equipped receiver, having the form of~\eqref{eq1}. After their MRC processing, the output signal of the receiver combiner is given as follows:
\beq
z_{\text{FAS}} \triangleq \sum_{k=1}^K \frac{h_{[k]}^* r_{[k]}}
{\left \lVert \undb{h}_{[K]} \right \rVert}
= \left \lVert \undb{h}_{[K]} \right \rVert s + \tilde{n}_{[K]} \, ,
\label{eq8}
\eeq
where $\tilde{n}_{[K]} \triangleq \sum_{k=1}^K h_{[k]}^* n_{[k]}/\left\lVert \undb{h}_{[K]} \right\rVert$, which, conditioned on $\undb{h}_{[K]}$, follows a complex Gaussian distribution as $\tilde{n}_{[K]} \big|_{\undb{h}_{[K]}} \sim {\mathcal{CN}} \left( 0 , \sigma_n^2 \right)$. Capitalizing on these statistics, the optimal ML receiver structure extracts the decoded symbol as follows:
\beq
\hat{s} \triangleq \arg \max_{s}
f \left( z_{\text{FAS}} \big| \undb{h}_{[K]} , s \right)
= \arg \min_{s} \left| z_{\text{FAS}}
- \left \lVert \undb{h}_{[K]} \right \rVert s \right|^2 ,
\label{eq9}
\eeq
where $f \left(\cdot\right)$ denotes the conditional probability density function of $z_{\text{FAS}}$ conditioned on $\undb{h}_{[K]}$ and $s$. For the transmission of real-valued $M$-ASK symbols, the receiver structure in \eqref{eq7} can be further simplified to:
\beq
\hat{s} = \arg \max_s \ s \, \Re \left\{ z_{\text{FAS}} \right\}
- \frac{s^2}{2} \left \lVert \undb{h}_{[K]} \right \rVert .
\label{eq10}
\eeq
For the transmission of $M$-PSK symbols, the optimal ML receiver structure simplifies to
\beq
\hat{s} = \arg \max_s \ \Re \left\{ z_{\text{FAS}} s^* \right\} .
\label{eq11}
\eeq
While \eqref{eq7} remains the most simplified expression for the ML receiver for $M$-QAM transmission, the receiver structure is modified for BFSK symbol transmission as:
\vspace{-0.2cm}
\beq
\Re \left\{ z_{\text{FAS}} \right\}
\begin{array}{c} A \vspace{-0.15cm} \\ > \vspace{-0.2cm} \\
< \vspace{-0.08cm} \\ \jmath A \end{array}
\Im \left\{ z_{\text{FAS}} \right\} .
\label{eq12}
\vspace{-0.2cm}
\eeq
The receiver structures \eqref{eq9}-\eqref{eq12} are used in the subsequent section to derive expressions for the SEP of the FA system.
\section{SEP Performance Analysis}
In this section, the performance of the considered FA systems is analyzed in terms of their SEPs for the various considered choices of one- and two-dimensional modulation schemes.
\subsection{$M$-ASK Transmission}
Let us denote the probability of a correct decision, given that the symbol $s_{m}$ ($m=1,\ldots,M$) from the set of $M$-ASK constellation is transmitted, by $P_{c_{m}}$. From the underlying decision rule in expression~\eqref{eq10}, we get that:
\bsub
\begin{align}
\! \! \! \! \! P_{c_1}
& \triangleq \Pr \left( s_1 \Re \left\{ \left \lVert \undb{h}_{[K]} \right \rVert
s_1 + \tilde{n}_{[K]} \right\} - \frac{s_1^2}{2}
\left \lVert \undb{h}_{[K]} \right \rVert \right. \nn \\
& \qquad \qquad 
\left. > s_2 \Re \left\{ \left \lVert \undb{h}_{[K]} \right \rVert
s_1 + \tilde{n}_{[K]} \right\} - \frac{s_2^2}{2}
\left \lVert \undb{h}_{[K]} \right \rVert \right) \nn \\
& = \Pr \left( \Re \left\{ \tilde{n}_{[K]} \right\}
< \left(s_2 - s_1 \right)
\frac{\left \lVert \undb{h}_{[K]} \right \rVert}{2} \right).
\label{eq13a}
\end{align}
Similarly, for $m=2,\ldots,M-1$, we have that:
\begin{align}
\! \! \! \! \! P_{c_{m}}
& \triangleq \Pr \left( \Re \left\{ \tilde{n}_{[K]} \right\}
< \left(s_{m+1} - s_{m} \right)
\frac{\left \lVert \undb{h}_{[K]} \right \rVert}{2} \right) \nn \\
& - \Pr \left( \Re \left\{ \tilde{n}_{[K]} \right\}
< - \left(s_{m} - s_{m-1} \right)
\frac{\left \lVert \undb{h}_{[K]} \right \rVert}{2} \right), 
\label{eq13b}
\end{align}
and
\beq
P_{c_M} \triangleq 1 - \Pr \left( \Re \left\{ \tilde{n}_{[K]} \right\}
< - \left(s_{M} - s_{M-1} \right)
\frac{\left \lVert \undb{h}_{[K]} \right \rVert}{2} \right).
\label{eq13c}
\eeq
\esub
From the statistics of the noise, it holds that $\Re \left\{ \tilde{n}_{[K]} \right\} \big|_{\undb{h}_{[K]}} \sim {\mathcal{N}} \left(0, \frac{\sigma_n^2}{2} \right)$. Therefore, the conditional probability of correct decisions, when conditioned on $\undb{h}_{[K]}$, is obtained as follows:
\begin{align}
& \! \! \! \!
P_{c_1 \big| \undb{h}_{[K]}} = P_{c_M \big| \undb{h}_{[K]}}
\triangleq 1 - Q \left( \sqrt{\frac{6 \gamma_{\text{FAS}}}{\left(M^2-1 \right)}} \right), \nn \\
& \! \! \! \! P_{c_m \big| \undb{h}_{[K]}}
\triangleq 1 - 2 Q \left( \sqrt{\frac{6 \gamma_{\text{FAS}}}{\left(M^2-1 \right)}} \right), \,\forall
m = 2, \ldots, M-1.
\label{eq14}
\end{align}
Thus, the conditional SEP, denoted by $P_{e \big| \undb{h}_{[K]}}^{\text{ASK}}$, is given by
\begin{align}
P_{e \big| \undb{h}_{[K]}}^{\text{ASK}} \! \!
& \triangleq \frac{1}{M} \sum_{m=1}^M P_{c_m \big| \undb{h}_{[K]}}
\! \! = \frac{2 \left(M-1 \right)}{M}
Q \left( \sqrt{\frac{6 \gamma_{\text{FAS}}}{\left(M^2-1 \right)}} \right) \nn \\
& \stackrel{(a)}{=} \frac{2 \left(M-1 \right)}{M \pi}
\! \! \int_0^{\pi/2} \! \! \! \! \! \!
\exp \left\{ - \frac{3 \gamma_{\text{FAS}}}
{\left(M^2-1 \right)\sin^2 \theta} \right\} \text{d} \theta , \! \!
\label{eq15}
\end{align}
where step $(a)$ arises from Craig's formula for the Gaussian $Q$-function. Unconditioning expression in \eqref{eq15}, gives an analytical expression for the SEP performance of the considered FA system for the case of $M$-ASK modulation:
\begin{align}
P_e^{\text{ASK}} & \triangleq \frac{2 \left(M-1 \right)}{M \pi}
\mathbb{E}_{\undb{h}_{[K]}}
\left[ \int_0^{\pi/2} \! \! \! \! \! \!
\exp \left\{ - \frac{3 \gamma_{\text{FAS}}}
{\left(M^2-1 \right) \sin^2 \theta} \right\} \text{d} \theta \right] \nn \\
& = \frac{2 \left(M-1 \right)}{M \pi}
\int_0^{\pi/2} \! \! \! \! \! \!
\Psi_{\gamma_{\text{FAS}}} \left( - \frac{3}
{\left(M^2-1 \right) \sin^2 \theta} \right)
\text{d} \theta .
\label{eq16}
\end{align}
Using the results presented in Appendix~A, an exact closed-form expression for the SEP in \eqref{eq16} is obtained as:
\beq
P_e^{\text{ASK}} = \frac{2 \left(M-1 \right)}{M}
{\mathcal J} \left( \frac{3}{\left(M^2-1 \right)}, \frac{\pi}{2} ; \Gamma_{\text{av}} \right).
\label{eq17}
\eeq
\subsection{$M$-PSK Transmission}
For the transmission of $M$-PSK symbols, the SEP when conditioned on $\undb{h}_{[K]}$ can be expressed as follows~\cite{10400440}:
\beq
P_{e \big| \undb{h}_{[K]}}^{\text{PSK}} \triangleq \frac{1}{\pi}
\int_{0}^{\frac{\pi \left(M-1 \right)}{M}}
\exp \left\{ - \frac{\gamma_{\text{FAS}}
\sin^2 \left( \frac{\pi}{M} \right)}
{\sin^2 \theta} \right\} \text{d} \theta.
\label{eq18}
\eeq
Unconditioning with respect to $\undb{h}_{[K]}$ and utilizing the results in Appendix A, the SEP of the considered FA system for the case of $M$-PSK modulation is obtained in closed form as:
\begin{align}
P_e^{\text{PSK}} & \triangleq \frac{1}{\pi}
\int_{0}^{\frac{\pi \left(M-1 \right)}{M}}
\Psi_{\gamma_{\text{FAS}}}
\left( - \frac{\sin^2 \left( \frac{\pi}{M} \right)}
{\sin^2 \theta} \right) \text{d} \theta \nn \\
& = {\mathcal{J}} \left( \sin^2 \left( \frac{\pi}{M} \right) ,
\frac{\pi \left(M-1 \right)}{M} ; \Gamma_{\text{av}} \right).
\label{eq19}
\end{align}
\subsection{$M$-QAM Transmission}
When the transmitter employs $M$-QAM for data modulation, the SEP of the FA system is expressed as follows~\cite{10400440}:
\begin{small}
\begin{align}
P_e^{\text{QAM}} 
& \triangleq \frac{4}{\pi} \left( 1 - \frac{1}{\sqrt{M}} \right)
\int_0^{\pi/2} \Psi_{\gamma_{\text{FAS}}}
\left( - \frac{3}{2 \left(M-1 \right) \sin^2 \theta} \right)
\text{d} \theta \nn \\
& - \frac{4}{\pi} \left( 1 - \frac{1}{\sqrt{M}} \right)^2
\int_0^{\pi/4} \Psi_{\gamma_{\text{FAS}}}
\left( - \frac{3}{2 \left(M-1 \right) \sin^2 \theta} \right)
\text{d} \theta .
\label{eq20}
\end{align}
\end{small}
Again, using the result in Appendix A, the SEP becomes:
\begin{align}
P_e^{\text{QAM}} & = 4 \left( 1 - \frac{1}{\sqrt{M}} \right)
{\mathcal{J}} \left( \frac{3}{2\left(M-1 \right)} ,
\frac{\pi}{2} ; \Gamma_{\text{av}} \right) \nn \\
& - 4 \left( 1 - \frac{1}{\sqrt{M}} \right)^2
{\mathcal{J}} \left( \frac{3}{2\left(M-1 \right)} ,
\frac{\pi}{4} ; \Gamma_{\text{av}} \right) \, .
\label{eq21}
\end{align}
\begin{figure*}[!t]
\centering
\subfigure[SEP versus $\Gamma_{\text{av}}$.]
{\includegraphics[height=1.6in,width=0.315\textwidth]{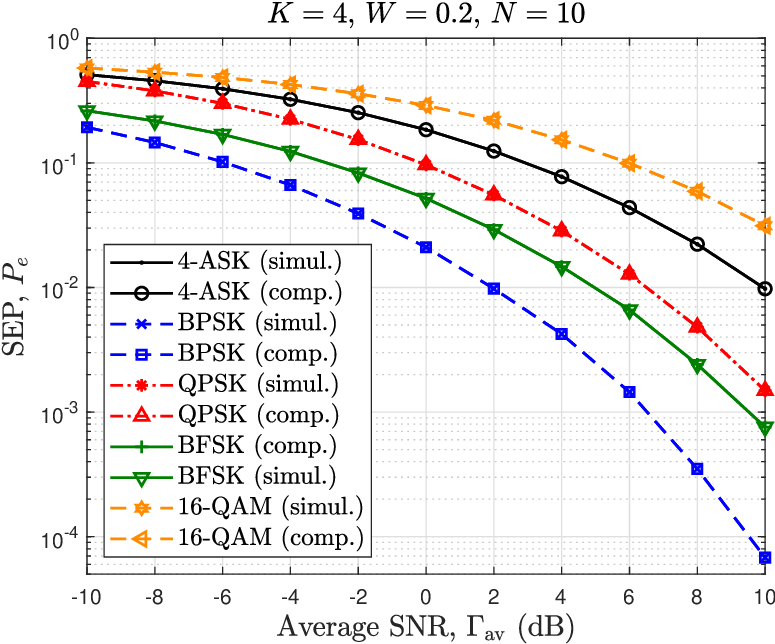}
\label{f1a}}
\subfigure[SEP versus $K$.]
{\includegraphics[height=1.6in, width=0.315\textwidth]{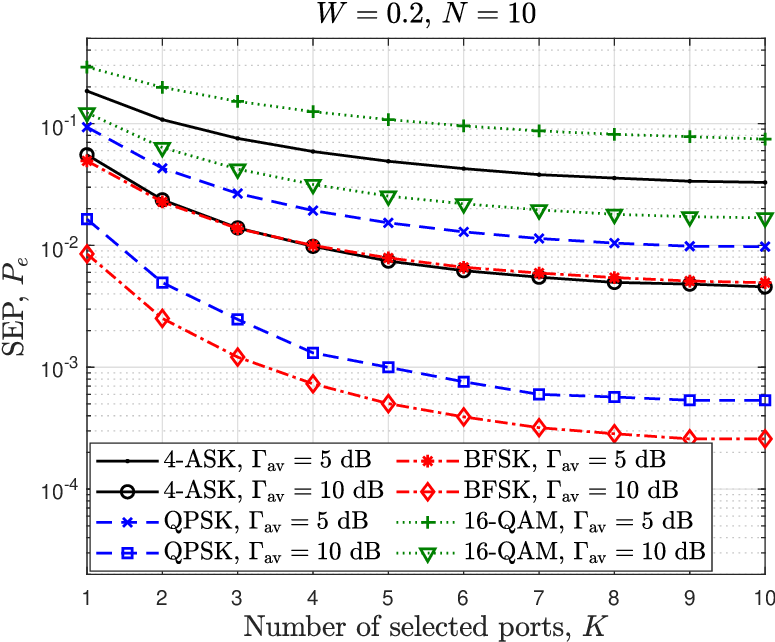}
\label{f1b}}
\subfigure[SEP versus $W$.]
{\includegraphics[height=1.6in,width=0.315\textwidth]{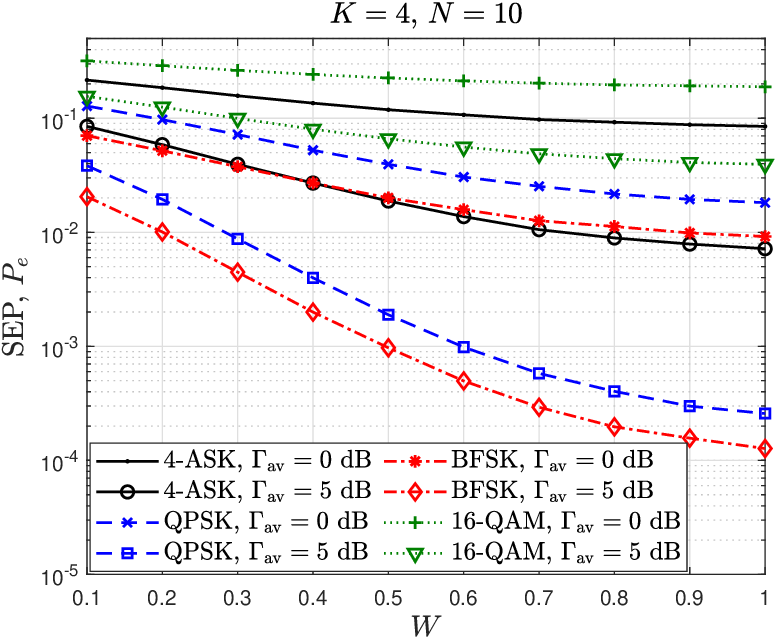}
\label{f1c}}
\caption{SEP performance for various modulation schemes with $N=10$ versus: (a) the average SNR of the FA system, $\Gamma_{\text{av}}$, for $K=4, W=0.2$; (b) the number of selected ports, $K$, for $W=0.2$, $\Gamma_{\text{av}}=5,10$ dB; and (c) $W$ for $K=4$, $\Gamma_{\text{av}}=0,5$ dB.}
\label{f1}
\vspace{-0.5cm}
\end{figure*}

\subsection{BFSK Transmission}
Using the receiver structure in~\eqref{eq12} for the case of BFSK signaling, the SEP of the considered FA system is given as:
\begin{align}
P_e^{\text{BFSK}} & \triangleq \frac{1}{2}
\Pr \left( \Re \left\{ z_{\text{FAS}} \right\}
> \Im \left\{ z_{\text{FAS}} \right\} \Big| s = \jmath A \right) \nn \\
& + \frac{1}{2} \Pr \left( \Re \left\{ z_{\text{FAS}} \right\}
< \Im \left\{ z_{\text{FAS}} \right\} \Big| s = A \right) \nn \\
& = \frac{1}{2} \Pr \left( \Re \left\{ \tilde{n}_{[K]} \right\}
- \Im \left\{ \tilde{n}_{[K]} \right\}
> A \left \lVert \undb{h}_{[K]} \right \rVert \right) \nn \\
& + \frac{1}{2} \Pr \left( \Re \left\{ \tilde{n}_{[K]} \right\}
- \Im \left\{ \tilde{n}_{[K]} \right\}
< - A \left \lVert \undb{h}_{[K]} \right \rVert \right).
\label{eq22}
\end{align}
Owing to the statistics of $\tilde{n}_{[K]}$, we have $\Re \left\{ \tilde{n}_{[K]} \right\} - \Im \left\{ \tilde{n}_{[K]} \right\} \big|_{\undb{h}_{[K]}} \sim {\mathcal{N}} \left(0, \sigma_n^2 \right)$. This results in the SEP expression conditioned on $\undb{h}_{[K]}$ to be obtained as follows:
\beq
P_{e \big| \undb{h}_{[K]}}^{\text{BFSK}}
\triangleq Q \left( \sqrt{\gamma_{\text{FAS}}} \right)
= \frac{1}{\pi} \int_0^{\pi/2}
\exp \left\{ - \frac{\gamma_{\text{FAS}}}
{2 \sin^2 \theta} \right\} \text{d} \theta .
\label{eq23}
\eeq
Unconditioning \eqref{eq23} with respect to $\undb{h}_{[K]}$ followed by utilizing the result of Appendix A, yields the exact closed-form expression for the system's SEP performance:
\beqarr
\! \! P_e^{\text{BFSK}}
\! = \frac{1}{\pi} \int_0^{\frac{\pi}{2}} \! \! \! \!
\Psi_{\gamma_{\text{FAS}}} \left( \! - \frac{1}{2 \sin^2 \theta}\right)
\text{d} \theta
= {\mathcal{J}} \left( \frac{1}{2} , \frac{\pi}{2} ; \Gamma_{\text{av}} \right).
\label{eq24}
\eeqarr
\subsection{Asymptotic SEP for $\Gamma_{\text{av}} \gg 1$}
For the asymptotic case of high average received SNR, i.e., $\Gamma_{\text{av}} \gg 1$, an asymptotic result for $\mathcal{J} \left( 0, \Theta ; \Gamma_{\text{av}} \right)$ is derived in \eqref{eqb2} in Appendix~B. Using \eqref{eqb2}, \eqref{eqb3a}, and \eqref{eqb3b}, analytical asymptotic expressions for SEP in this high SNR regime for all considered modulation schemes are  obtained as follows:
\bsub
\beq
P_{e \big| \Gamma_{\text{av}} \gg 1}^{\text{ASK}}
\approx \frac{\left(M-1 \right)^{N+1} \left(M+1 \right)^N \left(2N \right)!}
{2^{2N} 3^N M N! {\mathcal{K}} \left( K, \mu \right)
\Gamma_{\text{av}}^N} ,
\label{eq25a}
\eeq
\beq
P_{e \big| \Gamma_{\text{av}} \gg 1}^{\text{PSK}}
\approx \frac{N! \int_{0}^{\frac{\pi \left(M-1 \right)}{M}}
\sin^{2N} \theta \, \text{d} \theta}
{\pi \sin^{2N} \left( \frac{\pi}{M} \right)
{\mathcal{K}} \left( K, \mu \right) \Gamma_{\text{av}}^N}  ,
\label{eq25b}
\eeq
\begin{align}
& \! \! \! P_{e \big| \Gamma_{\text{av}} \gg 1}^{\text{QAM}}
\approx \left(1 - \frac{1}{\sqrt{M}} \right)
\frac{2^{N+1} \left(M-1 \right)^N N!}
{\pi 3^N {\mathcal{K}} \left( K, \mu \right) \Gamma_{\text{av}}^N} \nn \\
& \! \! \times \left( \frac{\pi \left(2N \right)!}
{2^{2N} \left(N! \right)^2}
- \left( 1 - \frac{1}{\sqrt{M}} \right)
B_{1/2} \left(N + \frac{1}{2} , \frac{1}{2} \right) \right) ,
\label{eq25c}
\end{align}
and
\beq
P_{e \big| \Gamma_{\text{av}} \gg 1}^{\text{BFSK}}
\approx \frac{\left(2N \right)!}
{2^{N+1} N! {\mathcal{K}} \left( K, \mu \right)
\Gamma_{\text{av}}^N} .
\label{eq25d}
\eeq
\esub
It can be seen that the effect of selecting $K$ FA ports out of the $N$ available is completely captured by the function $\mathcal{K} \left( K, \mu \right)$, and that the FA system achieves a diversity order of $N$.
\section{Numerical Results}
The comparison of simulations (via Monte Carlo trials using the optimal ML receiver structure in \eqref{eq9}), denoted by 'simul.', and the numerical evaluation of the derived SEP formulas, denoted by 'comp.', are presented in Fig.~\ref{f1}. The exactness of the plots verifies the correctness of the analytical framework. As shown in Fig.~\ref{f1a}, all the SEP plots tend to run parallel with increasing SNR, implying the same diversity order for the FAS. Figure~\ref{f1b} presents the plots of the SEP versus the number of selected ports at the FA for various modulation schemes. It is observed that, although the SEP improves with increasing $K$, it tends to saturate at higher $K$, with the rate of improvement diminishing as the modulation order increases.

Similarly, the plots of the SEP versus $W$ for different modulation schemes for two different $\Gamma_{\text{av}}$ values are presented in Fig.~\ref{f1c}. As expected, the SEP performance improves with an increase in the value of $W$ (which implies that the inter-port spacing increases). However, the effect of increasing $W$ is more prominent than increasing $K$ to achieve a lower SEP, as evident from the slopes of the SEPs in the plots. Moreover, this effect is more prominent at lower orders of the modulation scheme employed for data transmission.
\section{Conclusion}
An FA system comprising a single-antenna transmitter employing two-sided $M$-ASK, $M$-PSK, $M$-QAM, or BFSK signaling for data modulation has been studied in this paper. The FA receiver combined the best $K$ out of the $N$ available ports using MRC, and then used an optimal ML detection rule for data demodulation. Using a c.f. approach, novel exact and high-SNR asymptotic closed-form expressions for the system's SEP performance were presented. Numerical results showcased a prominent effect of the FA length as compared to $K$ in the reduction of the SEP, and validated the analytically derived diversity order $N$ for the considered FA system.
\section*{Appendix A: Integration of the c.f. of $\gamma_{\text{FAS}}$}
We consider the following integral:
\beq
{\mathcal J} \left( c, \Theta ; \Gamma_{\text{av}} \right)
= \frac{1}{\pi} \int_{0}^{\Theta} \Psi_{\gamma_{\text{FAS}}}
\left(- \frac{c}{\sin^2 \theta} \right) \text{d} \theta .
\label{eqa1}
\eeq
Using the expression of the c.f. of $\gamma_{\text{FAS}}$ from \eqref{eq6}, the latter integral in can be re-written as follows:
\begin{small}
\begin{align}
& {\mathcal{J}} \left( c, \Theta, \Gamma_{\text{av}} \right)
= \left( \frac{1 - \mu^2}{1+\left(N-1 \right) \mu^2} \right)
\sum_{p=0}^{\infty} \left( \frac{\mu^2}
{1+\left(N-1 \right) \mu^2} \right)^p \nn \\
& \times \! \! \! \! \! \sum_{\left(\ell_1,\ldots, \ell_N \right)}
\sum_{\left( q_1 , \ldots, q_N \right)}
\prod\limits_{k=2}^N \frac{{\mathcal{F}}_{q_1,\ldots,q_N} \left( c, K, \Theta \right)}{k^{q_k}}
{\ell_k \! + \! \sum\limits_{j=1}^{k-1} \left(\ell_j - q_j \right) \choose \ell_k},
\label{eqa2}
\end{align}
\end{small}
\hspace{-0.12cm}where the summations over $\left(\ell_1,\ldots,\ell_N \right)$ and $\left( q_1,\ldots,q_N \right)$ are the same as in \eqref{eq6} and \eqref{eq7}, and
\begin{small}
\begin{align}
& \! \! \! \! \! \!
{\mathcal{F}}_{q_1,\ldots,q_N} \left( c, K, \Theta \right)
\triangleq \frac{1}{\pi} \int_{0}^{\Theta} \! \! \!
\left( \frac{\sin^2 \theta}
{\sin^2 \theta + c \left(1-\mu^2 \right) \Gamma_{\text{av}}} \right)^{K+\sum\limits_{k=1}^K q_k} \nn
\end{align}
\end{small}
\vspace{-0.4cm}
\begin{small}
\begin{align}
& \qquad \quad \times \prod_{k=K+1}^N \left( \frac{\sin^2 \theta}
{\sin^2 \theta + \frac{c\left(1-\mu^2 \right) K \Gamma_{\text{av}}}{k}} \right)^{q_k+1} \text{d} \theta \nn \\
& = \frac{1}{\pi} \int_0^{\Theta} \prod_{k=1}^{\tilde{N}}
\left( \frac{\sin^2 \theta}{\sin^2 \theta
+ \frac{c \left(1-\mu^2 \right) K
\Gamma_{\text{av}}}{K+k-1}} \right)^{\eta_k} \text{d} \theta,
\label{eqa3}
\end{align}
\end{small}
with $\tilde{N}\triangleq N-K+1$ and $\eta_k$'s are given as follows:
\beq
\eta_k = \left\{ \begin{array}{ll}
\! \! q_1+\ldots+q_K +K \, , & \text{if } k=1 \\
\! \! q_{K+k-1} + 1 \, , & \text{if } k=2,\ldots, \tilde{N}
\end{array} \right..
\label{eqa4}
\eeq
Using the results from \cite[eq. (49)]{1025509}, it can be deduced:
\beq
{\mathcal{F}}_{q_1,\ldots,q_N} \left( c, K, \Theta \right)
= \sum_{k=1}^{\tilde{N}} \sum_{n=1}^{\eta_k}
\alpha_{k,n} \, \mathcal{G}_{k,n} \left( c_k , \Theta \right) ,
\label{eqa5}
\eeq
where $c_k \triangleq\left( \frac{c \left(1-\mu^2 \right) K \Gamma_{\text{av}}}{K+k-1} \right)^{-1}$ and
\begin{small}
\begin{align}
\alpha_{k,n} & \triangleq \prod_{\begin{array}{c}
\vspace{-0.2cm} {\scriptstyle p=1} \\ {\scriptstyle p \neq k}
\end{array}}^{\tilde{N}} \left(\frac{K+p-1}{p-k} \right)^{\eta_p}
\hspace{-1cm} \sum_{\begin{array}{c}
{\scriptstyle \left(\ell_1,\ldots, \ell_{\eta_k - n} \right)} \\
{\scriptstyle 0 \leq \ell_1, \ldots, \ell_{\eta_k - n} \leq \eta_k - n} \\
{\scriptstyle \ell_1 + 2\ell_2 + \ldots + \left( \eta_k - n \right) \ell_{\eta_k - n} = \eta_k - n}
\end{array}}
\hspace{-1cm} \prod_{q=1}^{\eta_k-n} \frac{1}{\ell_q!} \nn \\
& \hspace{1.5cm} \times \left( \frac{1}{q} \sum_{\begin{array}{c}
\vspace{-0.2cm} {\scriptstyle p=1} \\ {\scriptstyle p \neq k}
\end{array}}^{\tilde{N}} \eta_p
\left( \frac{K+k-1}{k-p} \right)^q \right)^{\ell_q} .
\label{eqa6}
\end{align}
\end{small}
Further, from \cite[eq. (51a)]{1025509}, the following holds:
\beq
\mathcal{G}_{k,n} \left( c_k , \Theta \right)
= \frac{\Theta}{\pi} + \sum_{i=1}^n \left(-1 \right)^i
{n \choose i} {\mathcal{H}}_{k,i} \left( c_k, \Theta \right) ,
\label{eqa7}
\eeq
where
\begin{small}
\begin{align}
& \! \! \! \! \! \!
{\mathcal{H}}_{k,i} \left( c_k, \Theta \right)
\triangleq \frac{1}{\pi \left( 1 + c_k \right)^{i-\frac{1}{2}}}
\sum_{l=0}^{i-1} {i-1 \choose l} {2l \choose l}
\left( \frac{c_k}{4} \right)^l \nn \\
& \quad \times \left[ \pi - \tan^{-1} \left( \sqrt{1+c_k}
\left| \tan \Theta \right| \right)
+ \sqrt{1+c_k} \frac{\tan \Theta}{2} \right. \nn \\
& \qquad \qquad \left. \times
\sum_{p=1}^l \frac{4^p}{{2p \choose p}}
\frac{1}{p \left(1+ \left(1+c_k \right) \tan^2 \Theta \right)^p} \right] .
\label{eqa8}
\end{align}
\end{small}
By using \eqref{eqa3}-\eqref{eqa8} in \eqref{eqa2} results in the solution to the integral of the c.f. of $\gamma_{\text{FAS}}$ shown in \eqref{eqa1}.
\section*{Appendix B: Asymptotic Expression for ${\mathcal J} \left( c, \Theta ; \Gamma_{\text{av}} \right)$}
For the case of high average SNR levels, i.e., for $\Gamma_{\text{av}} \gg 1$, the terms corresponding to $p=\ell_1=\ell_2=\ldots=\ell_N=0$ dominate in the expression of ${\mathcal J} \left( c, \Theta ; \Gamma_{\text{av}} \right)$ in \eqref{eqa2}. This results in $q_1=\ldots=q_N=0$, which further implies that:
\begin{small}
\begin{align}
& \! \! \! \! \! \! \! \! \! \! \! \! \! \! \! \! \! \!
{\mathcal{F}}_{0,\ldots,0}
\left(c, K, \Theta \right) \big|_{\Gamma_{\text{av}} \gg 1} \nn \\
& \approx \frac{1}{\pi} \int_0^{\Theta} \prod_{k=1}^{\tilde{N}}
\left( \frac{\left(K+k-1 \right) \sin^2 \theta}
{c \left(1-\mu^2 \right) K \Gamma_{\text{av}}} \right)^{\eta_k}
\text{d} \theta \nn \\
& \stackrel{(a)}{=}
\frac{N! \int_0^{\Theta} \sin^{2N} \theta \, \text{d} \theta}
{\pi \left(K-1 \right)! K^{\tilde{N}} c^N
\left(1-\mu^2 \right)^N \Gamma_{\text{av}}^N} ,
\label{eqb1}
\end{align}
\end{small}
\hspace{-0.11cm}where the step $(a)$ is based on the algebraic simplifications following $\sum_{k=1}^{\tilde{N}} \eta_k = N$ and $\prod_{k=1}^{\tilde{N}} \left(K+k-1 \right)^{\eta_k} = \frac{K^{K-1} N!}{\left(K-1 \right)!}$ for $q_1=\ldots=q_N=0$. This results in the asymptotic expression of ${\mathcal J} \left( c, \Theta ; \Gamma_{\text{av}} \right)$ for $\Gamma_{\text{av}} \gg 1$ to be obtained as:
\beq
{\mathcal J} \left( c, \Theta ; \Gamma_{\text{av}} \right) 
\big|_{\Gamma_{\text{av}} \gg 1}
\approx \frac{N! \int_0^{\Theta} \sin^{2N} \theta \, \text{d} \theta}
{\pi \mathcal{K} \left( K, \mu \right) c^N
\Gamma_{\text{av}}^N} ,
\label{eqb2}
\eeq
where
\bsub
\beq
{\mathcal K} \left( K, \mu \right)
\triangleq \left( 1 + \left(N-1 \right) \mu^2 \right)
\left(K-1 \right)! K^{\tilde{N}}
\left(1-\mu^2 \right)^{N-1} ,
\label{eqb3a}
\eeq
and
\begin{align}
& \! \! \! \!
\int_0^{\Theta} \sin^{2N} \theta \, \text{d} \theta
= \frac{\Theta}{2^{2N}} {2N \choose N} \nn \\
& \qquad + \frac{\left(-1 \right)^N}{2^{2N-1}}
\sum_{j=0}^{N-1} \left(-1 \right)^j {2N \choose j}
\frac{\sin \left( \left( 2N - 2j \right) \Theta \right)}
{\left(2N-2j \right)} .
\label{eqb3b}
\end{align}
\esub
\bibliographystyle{IEEEtran}
\bibliography{IEEEabrv,bibliography}

@ARTICLE{10400440,
  author={Basu, Aritra and others},
  journal={IEEE Trans. Wireless Commun.}, 
  title={Performance Analysis of {RIS}-Aided Index Modulation With Greedy Detection Over {Rician} Fading Channels}, 
  year={2024},
  volume={23},
  number={8},
  pages={8465-8479},
  doi={10.1109/TWC.2024.3350921}}

@ARTICLE{9264694,
   author={Wong, Kai-Kit and others},
  journal={IEEE Trans. Wireless Commun.}, 
  title={Fluid Antenna Systems}, 
  year={2021},
  volume={20},
  number={3},
  pages={1950--1962},
  doi={10.1109/TWC.2020.3037595}}

@ARTICLE{10379539,
  author={Wang, Zhe and others},
  journal={IEEE Commun. Surveys \&Tuts.}, 
  title={A Tutorial on Extremely Large-Scale {MIMO} for {6G}: Fundamentals, Signal Processing, and Applications}, 
  year={2024},
  volume={26},
  number={3},
  pages={1560--1605},
  doi={10.1109/COMST.2023.3349276}}

@ARTICLE{1025509,
  author={Mallik, R.K. and Win, M.Z.},
  journal={IEEE Trans. Commun.}, 
  title={Analysis of hybrid selection/maximal-ratio combining in correlated {Nakagami} fading}, 
  year={2002},
  volume={50},
  number={8},
  pages={1372-1383},
  doi={10.1109/TCOMM.2002.801495}}

@ARTICLE{11098630,
  author={Huangfu, Jiangsheng and others},
  journal={IEEE Trans. Wireless Commun.}, 
  title={Performance Analysis of Fluid Antenna System Under Spatially-Correlated {Rician} Fading Channels}, 
  year={2026},
  volume={25},
  number={},
  pages={1394-1407},
  doi={10.1109/TWC.2025.3590722}}

@INPROCEEDINGS{ganeshgc25,
  author={Tummi Ganesh and Soumya P. Dash and George C. Alexandropoulos},
  booktitle={Proc. IEEE GLOBECOM}, 
  title={Outage Probability Analysis of {RIS}-Assisted Fluid Antenna Systems over Double-{Nakagami}-$m$ Fading Channels}, 
  year={2025},
  address={Taipei, Taiwan},
  doi={10.1109/WCNC61545.2025.10978677}}

@ARTICLE{10375698,
  author={Lai, Xiazhi and others},
  journal={IEEE Commun. Lett.}, 
  title={On Performance of Fluid Antenna System Using Maximum Ratio Combining}, 
  year={2024},
  volume={28},
  number={2},
  pages={402-406},
  doi={10.1109/LCOMM.2023.3348028}}

@misc{ganesh2025OPMRC,
      title={Outage Probability Analysis of {MRC}-Based Fluid Antenna Systems under {Rician} Fading}, 
      author={Tummi Ganesh and Soumya P. Dash and Italo Atzeni},
      year={2025},
      eprint={2511.09474},
      archivePrefix={arXiv},
      primaryClass={eess.SP},
      url={https://arxiv.org/abs/2511.09474.}, 
}

@ARTICLE{10960696,
  author={Li, Yijia and others},
  journal={IEEE Wireless Commun. Lett.}, 
  title={Statistical {CSI}-Based Weighted Sum-Rate Maximization for Fluid Antenna-Aided Multiuser Communications}, 
  year={2025},
  volume={14},
  number={7},
  pages={1944-1948},
  doi={10.1109/LWC.2025.3559517}}

@ARTICLE{10934056,
  author={Chen, Jung-Chieh and others},
  journal={IEEE Wireless Commun. Lett.}, 
  title={Improved Joint Transmit and Receive Port Selection for Capacity Maximization in Fluid-{MIMO} Systems}, 
  year={2025},
  volume={14},
  number={6},
  pages={1693-1697},
  doi={10.1109/LWC.2025.3552939}}

@ARTICLE{10303274,
  author={New, Wee Kiat and others},
  journal={IEEE Trans. Wireless Commun.}, 
  title={An Information-Theoretic Characterization of {MIMO-FAS}: Optimization, Diversity-Multiplexing Tradeoff and $q$-Outage Capacity}, 
  year={2024},
  volume={23},
  number={6},
  pages={5541-5556},
  doi={10.1109/TWC.2023.3327063}}

@ARTICLE{10694739,
  author={Rostami Ghadi, Farshad and others},
  journal={IEEE Trans. Wireless Commun.}, 
  title={Physical Layer Security Over Fluid Antenna Systems: Secrecy Performance Analysis}, 
  year={2024},
  volume={23},
  number={12},
  pages={18201-18213},
  doi={10.1109/TWC.2024.3463488}}

@ARTICLE{10468625,
  author={Vega-Sánchez, José David and others},
  journal={IEEE Trans. Veh. Technol.}, 
  title={Fluid Antenna System: Secrecy Outage Probability Analysis}, 
  year={2024},
  volume={73},
  number={8},
  pages={11458-11469},
  doi={10.1109/TVT.2024.3376475}}

@INPROCEEDINGS{10978677,
  author={Ghadi, Farshad Rostami and others},
  booktitle={Proc. IEEE WCNC}, 
  title={Secrecy Performance Analysis of {RIS}-Aided Fluid Antenna Systems}, 
  year={2025},
  address={Milan, Italy},
  doi={10.1109/WCNC61545.2025.10978677}}

@ARTICLE{11360610,
  author={Mallik, Ranjan K. and Murch, Ross},
  journal={IEEE Trans. Wireless Commun.}, 
  title={Signaling for a Fluid Antenna System With Uniform Correlation}, 
  year={2026},
  volume={25},
  number={},
  pages={11223-11236},
  doi={10.1109/TWC.2026.3653914}}
\end{document}